\documentclass[preprint,authoryear,11pt,letterpaper]{elsarticle}
\usepackage[top=0.75in, bottom=0.75in, left=0.75in, right=0.75in]{geometry}
\usepackage{amsmath,amsthm}
\usepackage{amssymb}
\usepackage{amsfonts}
\usepackage{graphicx}
\usepackage{subfigure}
\usepackage{algorithm}
\usepackage{algorithmic}
\usepackage{tabularx,booktabs}
\usepackage{wrapfig}
\usepackage{tablefootnote}

\usepackage{xcolor}
\definecolor{darkblue}{rgb}{0.0,0.5,0.5}
\definecolor{blue}{rgb}{0.0,0.5,0.68}
\usepackage[colorlinks]{hyperref}
\hypersetup{colorlinks,breaklinks,linkcolor=darkblue,urlcolor=darkblue,anchorcolor=darkblue,citecolor=darkblue}
\usepackage{booktabs,caption}
\usepackage{multirow}
\usepackage{lscape}
\usepackage{epstopdf}
\usepackage{lineno}
\usepackage{subfigure}
\usepackage{multirow}
\usepackage{makecell}

\usepackage{microtype}
\usepackage{lineno}
\journal{Arxiv}

\begin{document}
\begin{frontmatter}
\title{Bayesian calibration of traffic flow fundamental diagrams using Gaussian processes}

\author[label1,label2]{Zhanhong Cheng}
\ead{zhanhong.cheng@mail.mcgill.ca}

\author[label1,label2]{Xudong Wang}
\ead{xudong.wang2@mail.mcgill.ca}

\author[label3]{Xinyuan Chen}
\ead{xinyuan.chen@nuaa.edu.cn}

\author[label2,label4]{Martin Trépanier}
\ead{mtrepanier@polymtl.ca}

\author[label1,label2]{Lijun Sun\corref{cor1}}
\ead{lijun.sun@mcgill.ca}

\address[label1]{Department of Civil Engineering, McGill University, Montreal, QC H3A 0C3, Canada}
\address[label2]{Interuniversity Research Centre on Enterprise Networks, Logistics and Transportation (CIRRELT), Montreal, QC H3T 1J4, Canada}
\address[label3]{College of Civil Aviation, Nanjing University of Aeronautics and Astronautics, Nanjing, China}
\address[label4]{Department of Mathematics and Industrial Engineering, Polytechnique Montreal, Montreal, QC H3T 1J4, Canada}
\cortext[cor1]{Corresponding author. Address: 492-817 Sherbrooke Street West, Macdonald Engineering Building, Montreal, Quebec H3A 0C3, Canada}


\begin{abstract}
Modeling the relationship between vehicle speed and density on the road is a fundamental problem in traffic flow theory. Recent research found that using the least-squares (LS) method to calibrate single-regime speed-density models is biased because of the uneven distribution of samples. This paper explains the issue of the LS method from a statistical perspective: the biased calibration is caused by the correlations/dependencies in regression residuals. Based on this explanation, we propose a new calibration method for single-regime speed-density models by modeling the covariance of residuals via a zero-mean Gaussian Process (GP). Our approach can be viewed as a generalized least-squares (GLS) method with a specific covariance structure (i.e., kernel function) and is a generalization of the existing LS and the weighted least-squares (WLS) methods. Next, we use a sparse approximation to address the scalability issue of GPs and apply a Markov chain Monte Carlo (MCMC) sampling scheme to obtain the posterior distributions of the parameters for speed-density models and the hyperparameters (i.e., length scale and variance) of the GP kernel. Finally, we calibrate six well-known single-regime speed-density models with the proposed method. Results show that the proposed GP-based methods (1) significantly reduce the biases in the LS calibration, (2) achieve a similar effect as the WLS method, (3) can be used as a non-parametric speed-density model, and (4) provide a Bayesian solution to estimate posterior distributions of parameters and speed-density functions.
\end{abstract}

\end{frontmatter}

\section{Introduction}

{
\color{red}


}

The fundamental diagram is a mathematical representation of speed-density or flow-density relationships, which is one of the underpinnings of traffic flow theory. Numerous works have been dedicated to modeling the relationships
since the first fundamental diagram Greenshields model \citep{greenshields1935study} was proposed. Thanks to the simple but efficient formulas, the single-regime speed-density models with a few meaningful parameters \citep[e.g.,][]{greenberg1959analysis, underwood1960speed, drake1965statistical, newell1961nonlinear, wang2011logistic} have been widely used in various applications, such as determining the road capacity \citep{wu2009derivation} and modeling traffic control strategies \citep{wang2014local}. However, as discussed in \citet{edie1961car,drake1965statistical}, single-regime models usually cannot perfectly fit the empirical data ranging from the light-traffic/free-flow regime to the congested/jam regime. Although multi-regime models (piecewise functions consisting of multiple single-regime models) can better fit empirical data, they lack the simplicity and the nice mathematical properties of single regime models (e.g., hard to determine the breakpoints \citep{kidando2020novel} and not differentiable at breakpoints).

\begin{figure}[!h]
    \begin{center}
    \includegraphics[width=.6\textwidth]{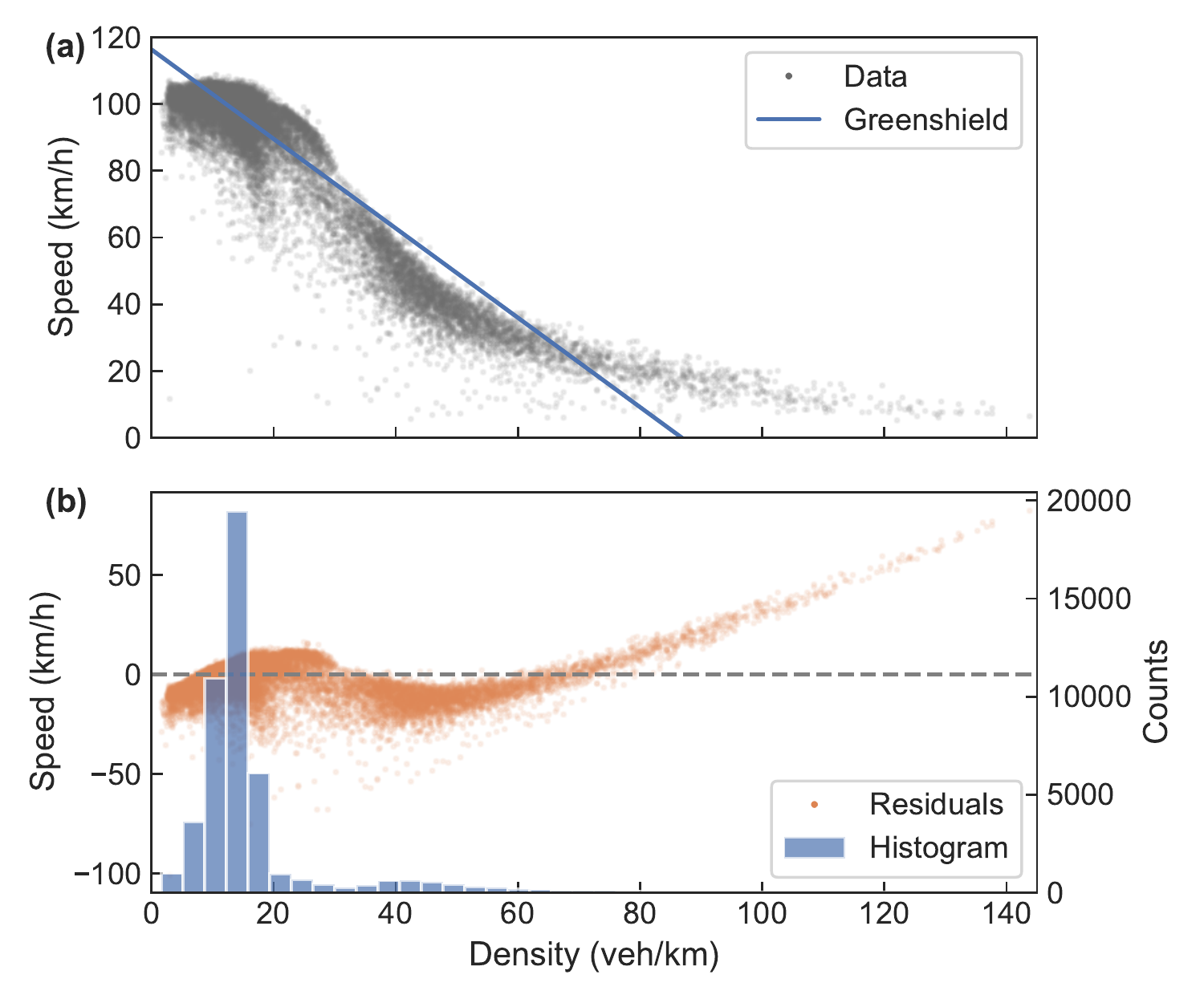}
    \caption{(a) Speed-density data and the least-squares fit of the Greenshields model; (b) Regression residuals and the data histogram in terms of density.}
    \label{fig:intro}
    \end{center}
\end{figure}

Besides the model/formula limitation, \citet{qu2015fundamental} pointed out that the calibration method, e.g., the least-squares (LS) method, can also cause inaccurate/biased parameter estimations for single-regime models due to the uneven distribution of speed-density observations. We illustrate this issue using a dataset collected by loop detectors from 76 stations on the Georgia State Route 400 (referred to as the GA400 dataset). The average speed, flow, and occupancy are collected for each station with a 20-second sampling interval and aggregated every 5 minutes. The GA400 dataset contains 47,815 observations and has been widely used in studying fundamental diagrams \citep{wang2011logistic, qu2015fundamental, qu2017stochastic, zhang2018reproducible, wang2021model}. Figure~\ref{fig:intro} (a) shows the speed-density observations of the GA400 dataset and a Greenshields model calibrated by the LS method; Figure~\ref{fig:intro} (b) shows the regression residuals and the data histogram in terms of density. Several issues can be found from the figure. (1) Skewed data: the distribution of the data is highly screwed (this is a general issue to most speed-density datasets); most observations have a small density (the densities of 86.8\% of the observations are in the range of being less than 20 veh/km). (2) Biased calibration: the Greenshields model calibrated by the LS method has much smaller absolute values of residuals for regions with more observations than regions with fewer observations. (3) Correlated residuals: there are strong correlations in the residuals.

To address the biased calibration issue, \citet{qu2015fundamental} proposed to calibrate single-regime speed-density models by a weighted least-squares (WLS) method. The weights in the WLS are determined by the distance in density between adjacent observations. Therefore, data from regions with fewer observations have larger weights to balance the biases. In a closely related work by \citet{zhang2018reproducible}, the authors developed a method to generate uniformly distributed samples from non-uniformly distributed observations and then calibrated the fundamental diagram by the LS method. This method has a similar performance to the WLS method. The above two works both attribute the biased calibration problem to the unevenly distributed data. However, we explain this issue from a statistical perspective and claim that the biased calibration of the LS is caused by the correlations in residuals. The LS method can be derived from the maximum likelihood estimation (MLE), where a fundamental assumption is that residuals are independent and identically distributed (i.i.d.) Gaussian noise. The LS method becomes biased when there are strong correlations in the residuals, such as in Figure~\ref{fig:intro} (b).

Based on our new explanation, first, this paper proposes a new calibration method for single-regime speed-density models, which reduces the biases by modeling the covariance of residuals using a zero-mean Gaussian Process (GP). Our approach can be viewed as a generalized least-squares (GLS) method with a specific covariance structure and is a generalization of the existing LS and WLS methods. Next, we use a sparse approximation to address the scalability issue in estimating GPs and apply a Markov chain Monte Carlo (MCMC) sampling scheme to obtain the posterior distributions of parameters for speed-density models. Finally, we test our GP-based method by calibrating six well-known single-regime speed-density models on the GA400 dataset. Results show that the proposed methods (1) significantly reduce the biases in the LS calibration, (2) achieve a similar effect as the WLS method, (3) can be used as a non-parametric speed-density model, and (4) provide a Bayesian solution to estimate posterior distributions of parameters and speed-density functions. It is necessary to clarify that the posterior distribution of parameters/speed-density functions in this paper is different from the stochastic fundamental diagram models \citep{muralidharan2011probabilistic,jabari2013stochastic,qu2017stochastic,wang2021model}. The stochastic fundamental diagram aims to use a set of speed-density functions to capture the distribution (uncertainty) of data, while the posterior distribution in this paper captures the distribution (uncertainty) of the estimation for the speed-density function.

There are four major contributions of this paper. First, we provide a statistical explanation for the biases of the LS method in calibrating speed-density models. Second, we propose to use a GP to resolve the biased calibration problem, and our method unites the existing LS and WLS methods into a generalized framework. Third, we introduce two solutions (sparse GP and MCMC sampling) to estimate the proposed GP-based model, making our method can scalable to large datasets and providing posterior distributions for the estimation. Lastly, the proposed GP regression can be used as a new non-parametric speed-density model.

The remainder of the papers is organized as follows. Section~\ref{sec:method} is the methodology part for the GP-based calibration method for speed-density models; the theory, the connections with existing models, and two solutions for the GP will be introduced in this section. Section~\ref{sec:test} is about experiments, where we will calibrate six single-regime speed-density models by the GP-based methods and compare the results with the LS and the WLS methods. Section~\ref{sec:conclusions} is left for conclusions and discussions.

\section{Methodology}\label{sec:method}
Consider a list of $n$ observations for speed $\mathbf{v} = \left[v_1, v_2, \cdots, v_n\right]^\top$ and density $\mathbf{k}=\left[k_1, k_2, \cdots, k_n \right]^\top$. We assume these paired observations can be explained by an unknown function $f$ and a random noise term
\begin{equation}\label{eq:regression}
    v_i = f(k_i) + \varepsilon_i.
\end{equation}
Conventional approaches directly replace the unknown function with a given-form single-regime speed-density model $m(k)$ and calibrate model parameters by the LS method. However, most single-regime speed-density models are not a perfect fit for the speed-density relationship, and the residuals $\left(v-m(k)\right)$ are, therefore, not independent. The correlations in regression residuals violate a fundamental assumption---independent noise---in the LS estimation\footnote{The LS method can be derived from a maximum likelihood estimation when assuming i.i.d.\ Gaussian noise \citep[][Chapter 3]{bishop2006pattern}. Although, when the function is linear, according to the Gauss-Markov theorem, the LS estimation is unbiased when the noise is independent, zero-mean, and homoscedastic with finite variance (could be non-Gaussian). The noise should be independent in either case.}, which causes biases in parameter calibrations, particularly in the case of unevenly distributed observations \citep{qu2015fundamental}. Although the generalized least-squares (GLS) method \citep{aitken1936iv} can overcome the biases of the LS method, the estimation for the GLS in our case is difficult because of the unknown residual covariance matrix, non-linear speed-density function, and the large number of observations.

We propose a new calibration method by Gaussian Process (GP) regression to resolve the above issue. Section~\ref{sec:gp} elaborates on the foundation of the GP-based method and its connections with the existing WLS and LS methods. We further introduce a sparse GP method in Section~\ref{sec:sparse gp} to resolve the scalability issue in the GP regression. Finally, Section~\ref{sec:mcmc} introduces an MCMC sampling method for variational sparse GP \citep{hensman2015mcmc} to obtain parameters' posteriors in speed-density models.

\subsection{Gaussian Process regression}\label{sec:gp}
We impose a GP prior to the unknown function $f$:
\begin{align}
    f(k) & \sim\mathcal{GP}\left(m(k), c\left(k, k'\right)\right), \label{eq:gp}\\
    c(k, k') &= \sigma^2\exp\left(-\frac{(k-k')^2}{2 \ell^2}\right), \label{eq:se kernel}
\end{align}
where the mean function $m(k)$ is a speed-density model; the covariance function (a.k.a. kernel function) $c\left(k, k'\right)$ captures the covariance of the residuals. The function value at observed points follows a multivariate Gaussian distribution $p\left(\mathbf{f}\right) = \mathcal{N}\left(m(\mathbf{k}), \mathbf{C}_{nn}\right)$, where $\mathbf{C}_{nn}$ is a covariance matrix with elements $\mathbf{C}_{nn}\left[i,j\right] = c(k_i, k_j)$. The squared exponential (SE) kernel in Eq.~\eqref{eq:se kernel} is one of the most commonly used kernels in GP (although using other kernels is possible), characterized by smooth function values; the variance $\sigma^2$ and the length scale $\ell$ are hyperparameters that should be estimated from data. To better understand our method, an equivalent form for Eq.\eqref{eq:regression}--Eq.\eqref{eq:gp} is
\begin{align}
    v &= m(k) + g(k) + \varepsilon, \\
    g(k) &\sim \mathcal{GP}\left(0, c\left(k, k'\right)\right),
\end{align}
where $g(k)$ can be regarded as a systematic error component that cannot be modeled by the speed-density function $m(k)$, and we model the systematic error component by a zero-mean GP. To avoid ambiguity, we refer to $\left(v-m(k)\right)$ the regression residual and the $\varepsilon$ the noise thereafter.

Assuming an i.i.d.\ Gaussian noise term of $\varepsilon$ brings great convenience to calibrating a GP. In this case, the joint distribution of $n$ speed observations follows a multivariate Gaussian distribution $p(\mathbf{v}) = \mathcal{N}\left(m(\mathbf{k}), \mathbf{C}_{nn} + \sigma_{\varepsilon}^2 \mathbf{I} \right)$, where $\sigma_{\epsilon}^2$ is the variance of $\varepsilon$, and $\mathbf{I}$ is an identity matrix. We denote by $\boldsymbol{\beta}$ the parameters in the mean function and by $\boldsymbol{\theta} = \left\{\boldsymbol{\beta}, \ell, \sigma^2, \sigma_{\varepsilon}^2 \right\}$ the parameters of the GP. We can estimate model parameters by Maximum marginal Likelihood Estimation (MLE), which is equivalent to minimizing the following negative log marginal likelihood with respect to $\boldsymbol{\theta}$:
\begin{align}
    - \log p(\mathbf{v} | \boldsymbol{\theta}) &=  - \log \mathcal{N}\left(\mathbf{v} | m(\mathbf{k}), \mathbf{C}_{nn} + \sigma_{\varepsilon}^2 \mathbf{I} \right) \\
    & = \frac{1}{2}(\mathbf{v}-m(\mathbf{k}))^\top \left( \mathbf{C}_{nn} + \sigma_{\varepsilon}^2 \mathbf{I}\right)^{-1}(\mathbf{v}-m(\mathbf{k})) + \frac{1}{2}\log (|\mathbf{C}_{nn} + \sigma_{\varepsilon}^2 \mathbf{I}|) + \frac{n}{2}\ln (2\pi). \label{eq:neg log marginal likelihood}
\end{align}
This minimization can be solved numerically by gradient-based methods.

Indeed, the above GP-based calibration can be viewed as a generalized least-squares (GLS) method with a structured residual covariance matrix $\boldsymbol{\Sigma} = (\mathbf{C}_{nn}+\sigma_{\varepsilon}^2 \mathbf{I})$. The quadratic term $\frac{1}{2}(\mathbf{v}-m)^\top \boldsymbol{\Sigma}^{-1} (\mathbf{v}-m)$ in Eq.~\eqref{eq:neg log marginal likelihood} plays a role of minimizing the regression residuals in squared Mahalanobis length \citep{mahalanobis1936generalized}. When $\mathbf{C}_{nn}=\boldsymbol{0}$ and $\sigma_{\varepsilon}=1$, minimizing Eq.~\eqref{eq:neg log marginal likelihood} becomes an LS estimation. When $\mathbf{C}_{nn}=\boldsymbol{0}$ and allowing different $\sigma_{\varepsilon}$ along the diagonal of $\boldsymbol{\Sigma}$ (i.e., heteroscedasticity), Eq.~\eqref{eq:neg log marginal likelihood} is equivalent to the WLS method adopted by \citet{qu2015fundamental}. In this paper, we use the SE kernel in Eq.~\eqref{eq:se kernel} to construct $\mathbf{C}_{nn}$, accounting for the dependencies in the residuals. Meanwhile, the second term $\frac{1}{2}\log (|\boldsymbol{\Sigma}|)$ in Eq.~\eqref{eq:neg log marginal likelihood} penalizes the complexity of the covariance matrix, leaving spaces for the mean function to explain more variance of the data. Therefore, the GP-based calibration is a generalization of a wide range of calibration methods with different covariance structures. Besides, although the main purpose of this paper is to calibrate a speed-density model, the GP itself can function as a non-parametric model for data-driven fundamental diagrams.

\subsection{Sparse GP}\label{sec:sparse gp}
The GP regression introduced in Section~\ref{sec:gp} does not scale to large datasets because of the $\mathcal{O}(n^3)$ time complexity and the $\mathcal{O}(n^2)$ storage complexity. Fortunately, various scalable GP methods have been developed to overcome this difficulty \citep{liu2020gaussian}. We applied a sparse GP method using inducing points to address the computational issue.

The computational bottleneck of a GP is calculating the inverse and the determinant of the covariance matrix $\boldsymbol{\Sigma}$. The idea of the sparse GP is to approximate the covariance matrix using a small set of $u$ auxiliary inducing points at $\mathbf{k}_u$. The function values at these inducing points follow the same GP prior $p\left(\mathbf{f}_u\right) = \mathcal{N}\left(m(\mathbf{k}_u), \mathbf{C}_{uu}\right)$, and it assumes that the function values $\mathbf{f}$ at observed points and $\mathbf{f}_{*}$ at new locations are conditionally independent given $\mathbf{f}_u$, i.e., $p\left(\mathbf{f}_{*}| \mathbf{f}, \mathbf{f}_u\right)=p\left(\mathbf{f}_{*}| \mathbf{f}_u\right)$. Using inducing points, the function covariance matrix is approximated with a low-rank representation $\mathbf{C}_{nn} \approx \mathbf{C}_{nu} \mathbf{C}_{uu}^{-1} \mathbf{C}_{nu}^{\top}$, where $\mathbf{C}_{nu}\in\mathbb{R}^{n\times u}$ is the covariance matrix between $\mathbf{f}$ and $\mathbf{f}_u$. Next, the inverse and the determinant in Eq.~\eqref{eq:neg log marginal likelihood} can be simplified by the matrix inversion lemma (also known as the Sherman-Morrison-Woodbury formula) and the matrix determinant lemma, respectively; it follows
\begin{align}
    \left(\mathbf{C}_{nu}\mathbf{C}_{uu}^{-1}\mathbf{C}_{nu}^{\top}+\sigma_{\varepsilon}^2 \mathbf{I}\right)^{-1} &=
    \sigma_{\varepsilon}^{-2}\mathbf{I} -
    \sigma_{\varepsilon}^{-4}\mathbf{C}_{nu} \left(\sigma_{\varepsilon}^{-2}\mathbf{C}_{nu}^{\top}\mathbf{C}_{nu} + \mathbf{C}_{uu}\right)^{-1}
    \mathbf{C}_{nu}^{\top},\label{eq:woodbury} \\
    |\mathbf{C}_{nu}\mathbf{C}_{uu}^{-1}\mathbf{C}_{nu}^{\top}+\sigma_{\varepsilon}^2 \mathbf{I}| &= |\sigma_{\varepsilon}^{-2}\mathbf{C}_{nu}^{\top}\mathbf{C}_{nu} + \mathbf{C}_{uu}||\mathbf{C}_{uu}^{-1}|\sigma_{\varepsilon}^{2n}.\label{eq:determinant}
\end{align}
With Eq.~\eqref{eq:woodbury}--Eq.~\eqref{eq:determinant}, the time and storage complexity of estimating a sparse GP using the MLE reduces to $\mathcal{O}(nu^2)$.

Although there are methods for choosing the number and the locations for inducing points, we use evenly spaced twenty points between the minimum and the maximum observed density as inducing points. This simple choice for inducing points will suffice for our problem since the shape of the speed-density relationship is pretty simple. Besides, because this paper focuses on the mean function $m(k)$ rater than the function value $f(k)$, many other sparse GP methods in the literature will have the same estimation results for our problem. For example, the Nyström GP \citep{williams2000using}, the subset of regressors \citep{smola2000sparse}, and the deterministic training conditional approximation \citep{csato2002sparse, seeger2003fast}. Interested readers are referred to an article by \citet{quinonero2005unifying} for the details of these sparse GP models.

\subsection{MCMC for variational sparse GP}\label{sec:mcmc}
The estimation methods in Section~\ref{sec:gp} and Section~\ref{sec:sparse gp} only works when the noise term $\varepsilon$ follows a Gaussian distribution. However, as we can see from Figure~\ref{fig:intro} (a), there are a fair amount of outliers in the data and using a long-tailed distribution (e.g., Student-$t$ distribution) for the noise term is more robust in such a case. Moreover, traffic practitioners often have prior knowledge about model parameters (e.g., free flow speed and jam density); knowing the uncertainty (posterior distribution) of parameters is much more interesting than a point estimation. Therefore, a Bayesian estimation based on an MCMC sampling method developed by \citet{hensman2015mcmc} for variational sparse GP is introduced in this section to resolve the above issues.

The sparse GP in Section~\ref{sec:sparse gp} uses approximate GP prior (in the covariance matrix) but exact inference. In contrast, the variational sparse GP \citep{titsias2009variational,hensman2015mcmc} uses exact GP prior but approximate inference. Following the notations in Section~\ref{sec:sparse gp}, unknown values in a sparse GP include $\mathbf{f}$, $\mathbf{f}_u$, $\mathbf{f}_{*}$, and $\boldsymbol{\theta}$. The idea of the variational sparse GP by \citet{hensman2015mcmc} is to approximate the intractable true posterior $p\left(\mathbf{f}, \mathbf{f}_u, \mathbf{f}_{*}, \boldsymbol{\theta} | \mathbf{v} \right)$ by another distribution $q\left(\mathbf{f}, \mathbf{f}_u, \mathbf{f}_{*}, \boldsymbol{\theta} \right)$ by minimizing their Kullback-Leibler (KL) divergence $\mathrm{KL}\left(q\left(\mathbf{f}, \mathbf{f}_u, \mathbf{f}_{*}, \boldsymbol{\theta} \right) \parallel p\left(\mathbf{f}, \mathbf{f}_u, \mathbf{f}_{*}, \boldsymbol{\theta} | \mathbf{v} \right)\right)$. \citet{hensman2015mcmc} found that the minimum KL divergence is reached with an optimal variational distribution:
\begin{equation}
    \log \hat{q}\left( \mathbf{f}_u, \boldsymbol{\theta} \right) =\mathbb{E}_{p(\mathbf{f} \mid \mathbf{f}_u, \boldsymbol{\theta})} \left[ \log p\left( \mathbf{y}\mid \mathbf{f} \right) \right] +\log p\left( \mathbf{f}_u \mid \boldsymbol{\theta} \right) +\log p\left( \boldsymbol{\boldsymbol{\theta} } \right) -\log C, \label{eq:log_p_mcmc}
\end{equation}
where $p\left(\boldsymbol{\theta}\right)$ is the prior distribution of parameters, $C$ is a normalizing constant. The first term on the right side of Eq.~\eqref{eq:log_p_mcmc} is the expected log-likelihood for noise $\varepsilon$. This expectation can be factorized across data when assuming i.i.d.\ noise; next, this term can be analytically solved for Gaussian or Poisson noise, and can be approximated by Gauss-Hermite quadrature for other types of noise. The time complexity for computing Eq.~\eqref{eq:log_p_mcmc} is $\mathcal{O}\left(nu^2\right)$.

Direct sampling from Eq.~\eqref{eq:log_p_mcmc} using the MCMC is inefficient because the function value $\mathbf{f}_u$ and GP parameters $\boldsymbol{\theta}$ are usually strongly coupled \citep{murray2010slice}. Therefore, the following reparameterization (known as whitening the prior) is introduced to improve the sampling efficiency. Define an ancillary random variable $\boldsymbol{\nu}\sim\mathcal{N}\left(\mathbf{0}, \mathbf{I}\right)$, such that  $\mathbf{f}_u=\mathbf{R}\boldsymbol{\nu}$ with $\mathbf{R}\mathbf{R}^{\top}=\mathbf{C}_{uu}$. The optimal log probability density for $\boldsymbol{\nu}$ and $\boldsymbol{\theta}$ is thus
\begin{equation}
    \log \hat{q}\left( \boldsymbol{\nu},\boldsymbol{\boldsymbol{\theta} } \right) =\mathbb{E}_{p\left( \mathbf{f}\mid \mathbf{f}_u=\mathbf{R}\boldsymbol{\nu} \right)}\left[ \log p\left( \mathbf{y}\mid \mathbf{f} \right) \right] +\log p\left( \boldsymbol{\nu} \right) +\log p\left( \boldsymbol{\boldsymbol{\theta} } \right) -\log C. \label{eq:log_mcmc2}
\end{equation}
We apply No-U-Turn sampler \citep{hoffman2014no}, a type of Hamiltonian MCMC \citep{betancourt2017conceptual} method that automatically tunes step size, to efficiently draw samples from Eq.\eqref{eq:log_mcmc2}. Next, parameters' posterior distributions can be obtained from the samples.

\section{Tests}\label{sec:test}
This section tests the effect of using the GP in calibrating single regime speed-density models. Section~\ref{sec:settings} introduces the settings of the experiment and models. The calibration results of different models are compared in Section~\ref{sec:compare}. Section~\ref{sec:mcmc_results} will show the posterior distributions obtained by the MCMC solution for our GP method.


\subsection{Experiment settings}\label{sec:settings}
We calibrate the six speed-density models in Table~\ref{tab:single_model} on the GA400 dataset using the GP regression, and the calibration results are compared with the WLS \citep{qu2015fundamental} and the LS methods. We use the sparse approximation for the MLE because of the large number of observations (47,581 points). As mentioned in Section~\ref{sec:sparse gp}, twenty inducing points are evenly spaced from minimum to maximum observed densities.

\begin{table}[!htpb]
    \centering
    \caption{\label{tab:single_model} Single-regime speed-density models.}
    \begin{tabular}{p{0.48\textwidth}p{0.32\textwidth}p{0.1\textwidth}}
    \toprule
    Single-regime models           & Function & Parameters \\ \midrule
    Greenshields \citep{greenshields1935study}                   & $v =v_f(1-\frac{k}{k_j})$         & $v_f,~k_j$\\
    Greenberg \citep{greenberg1959analysis}                     & $v=v_0\ln{(\frac{k_j}{k})}$        & $v_0,~k_j$           \\
    Underwood \citep{underwood1960speed}                     & $v=v_f\exp{(-\frac{k}{k_0})}$        & $v_f,~k_0$           \\
    Northwestern \citep{drake1965statistical}                  &$v=v_f\exp{\left(-\frac{1}{2}(\frac{k}{k_0})^2\right)}$ & $v_f,~k_0$            \\
    Newell \citep{newell1961nonlinear}                       &$v=v_f\left(1-\exp{\left(-\frac{\lambda}{v_f}(\frac{1}{k} - \
    \frac{1}{k_j})\right)}\right)$         &   $v_f,~k_j,~\lambda$         \\
    Three-parameter logistic (3PL) \citep{wang2011logistic} & $v = \frac{v_f}{1+\exp{({k-k_c}/{\theta})}}$        & $v_f, ~k_c,~\theta$            \\ \bottomrule
    \end{tabular}
\end{table}

We also test using the MCMC sampling for variational sparse GP to obtain the posterior distributions of model parameters. To make the estimation more robust to the outliers in the data, we assume the noise term $\varepsilon$ follows an i.i.d.\ zero-mean Student-$t$ distribution. The choice of priors in the MCMC sampling are as follows:
\begin{itemize}
    \item Gaussian priors are used for all parameters regarding the speed-density functions. For a parameter $x$, its prior mean is set as the WLS estimation $x_{\mathrm{WLS}}$, and the standard deviation is set by $\max\left(\frac{x_{\mathrm{WLS}}}{6}, 10\right)$.
    \item Other hyperparameters are positive values, including $\sigma^2$ and $\ell$ in the kernel functions and scale and degree of freedom for the Student-$t$ noise. Therefore, we set their prior by Half-Cauchy distribution $p(x)=\frac{2}{\pi \left(1 + x^2\right)}, x\in[0, +\infty)$.
\end{itemize}

We use a Python package named GPflow \citep{matthews2017gpflow} to set up the variational sparse GP introduced in Section~\ref{sec:mcmc}. Next, the No-U-Turn sampling routine implemented in TensorFlow Probability \citep{dillon2017tensorflow} is used to draw samples from Eq.~\eqref{eq:log_mcmc2}. The first 2,000 steps are taken as the ``burn-in'' period to tune the step size and converge the sequence; we collect the next 3,000 steps to calculate posterior distributions.

\subsection{Comparison with other models}\label{sec:compare}
\begin{figure}[htpb]
    \begin{center}
    \includegraphics[width=0.8\textwidth]{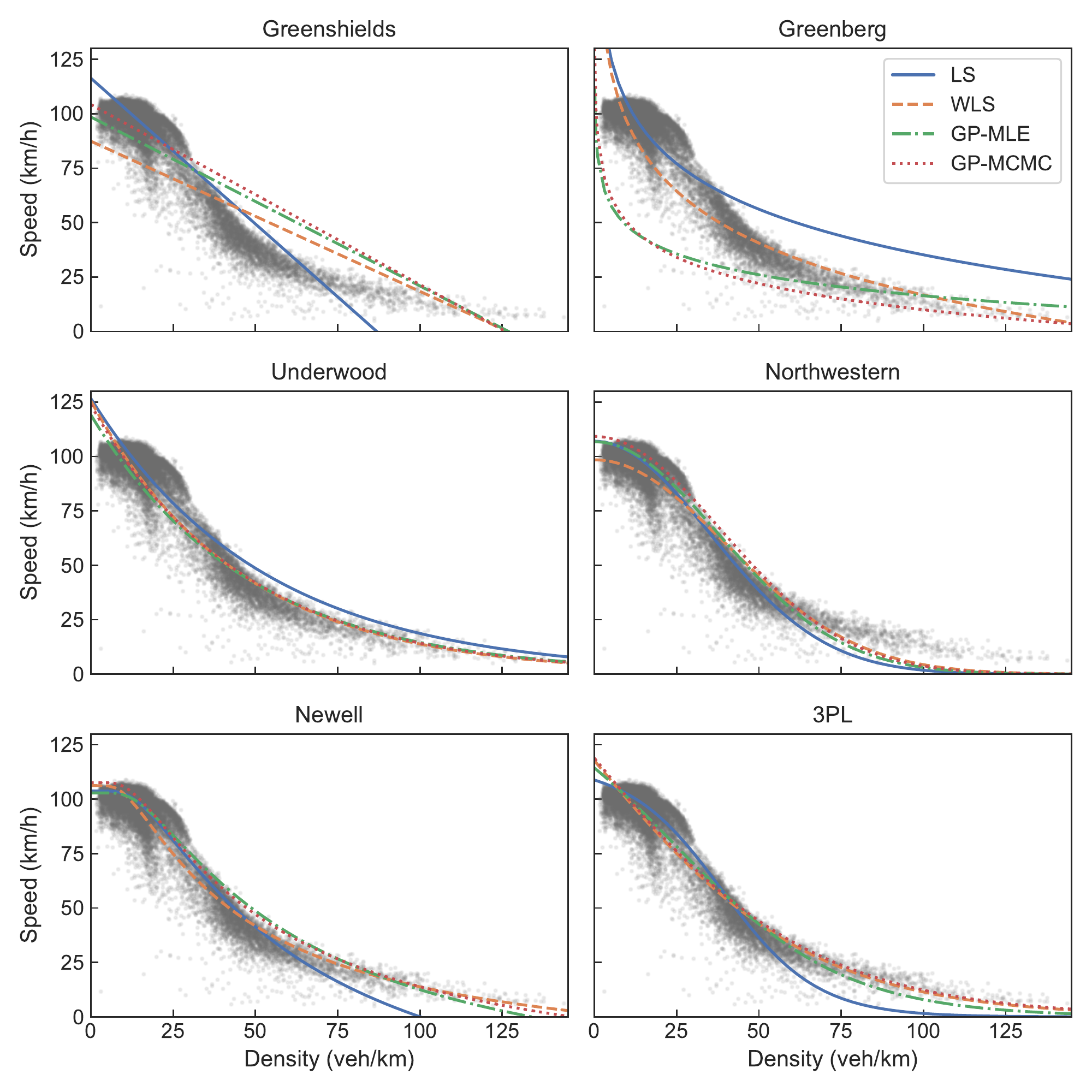}
    \caption{Speed-density models calibrated by LS, WLS, sparse GP by MLE, and the mean of MCMC sampling for variational sparse GP.}
    \label{fig:compare}
    \end{center}
\end{figure}

The six speed-density models calibrated by the four methods are compared in Figure~\ref{fig:compare}, where the result of the GP-MCMC is the average of posterior samples. We can see that the LS method results in speed-density models that fit the low-density region ($\leq 50$ veh/km) well but are inappropriate for the high-density region ($> 50$ veh/km). In contrast, the WLS and the two GP methods generally produce more ``unbiased'' speed-density models in the entire density range. Two GP methods have very similar calibration results to the WLS method for the Underwood, the Newell, and the 3PL speed-density models. For the Greenshields and the Northwestern speed-density model, the free flow speed ($v$ at $k=0$) estimated by GP methods is higher than which estimated by the WLS method.

An exception for the GP method is the Greenberg speed-density model, where the mean function does not fit the data well. This is caused by a limitation of the Greenberg model: $v\rightarrow+\infty$ when $k\rightarrow+0$. A Greenberg model that fits most observations well (e.g., the WLS estimation) has a significantly higher estimation for speed at the near zero-density area; this ``abrupt surge in the regression residual'' is very unlikely to happen in the GPs because the SE kernel that we use favors a smooth function. Therefore, a GP-based estimation produces a flat mean function $m(k)$ and a significant systematic error component $g(k)$ in the Greenberg model. The GP essentially uses a different way than the other methods to reveal the inappropriateness of the Greenberg model. Many measures can be taken to improve the GP estimation for the Greenberg model, inducing (1) adding a small positive shift parameter $k_s$ so $\lim_{k\to 0^+}v_0\ln\left(\frac{k_j}{k+k_s}\right)$ is not infinity, (2) fitting the inverse function $k=f^{-1}(v)$, (3) using other kernels in the GP (e.g., non-stationary kernels), and (4) discarding data with small densities.

\begin{figure}[!h]
    \begin{center}
    \includegraphics[width=0.8\textwidth]{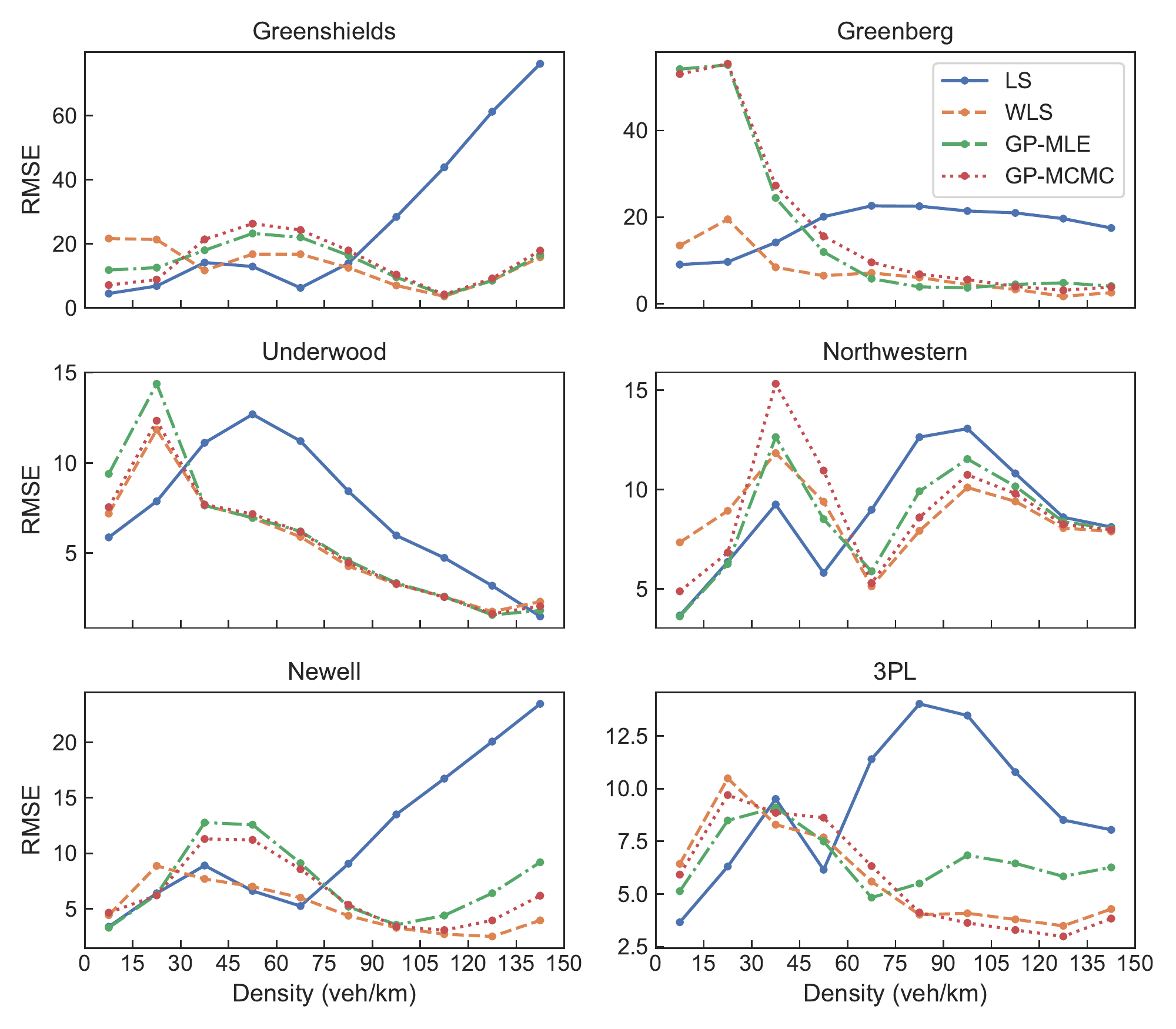}
    \caption{The RMSE of the four calibration methods on the six single-regime models at different density groups.}
    \label{fig:rmse}
    \end{center}
\end{figure}

In Figure~\ref{fig:rmse}, we divide data into ten groups by density from $k=0$ to $k=150$ at an interval of $15$ to better understand the fitting performance. We calculate the root-mean-square error (RMSE) for each density group between speed-density models and data. We can find from Figure~\ref{fig:rmse} that the WLS and the two GP methods alleviate the biased calibration of the LS method and greatly reduce the RMSE for high-density (around $k>75$) groups. Except for the Greenberg model, it is hard to conclude which one between the GP and WLS method has a better calibration result. Besides, the RMSE of the GP using the MCMC sampling is a little lower than the MLE methods for the Newell and the 3PL models, which could be a benefit from assuming the Student-$t$ noise and using the priors.

\begin{figure}[!h]
\begin{center}
\includegraphics[width=0.5\textwidth]{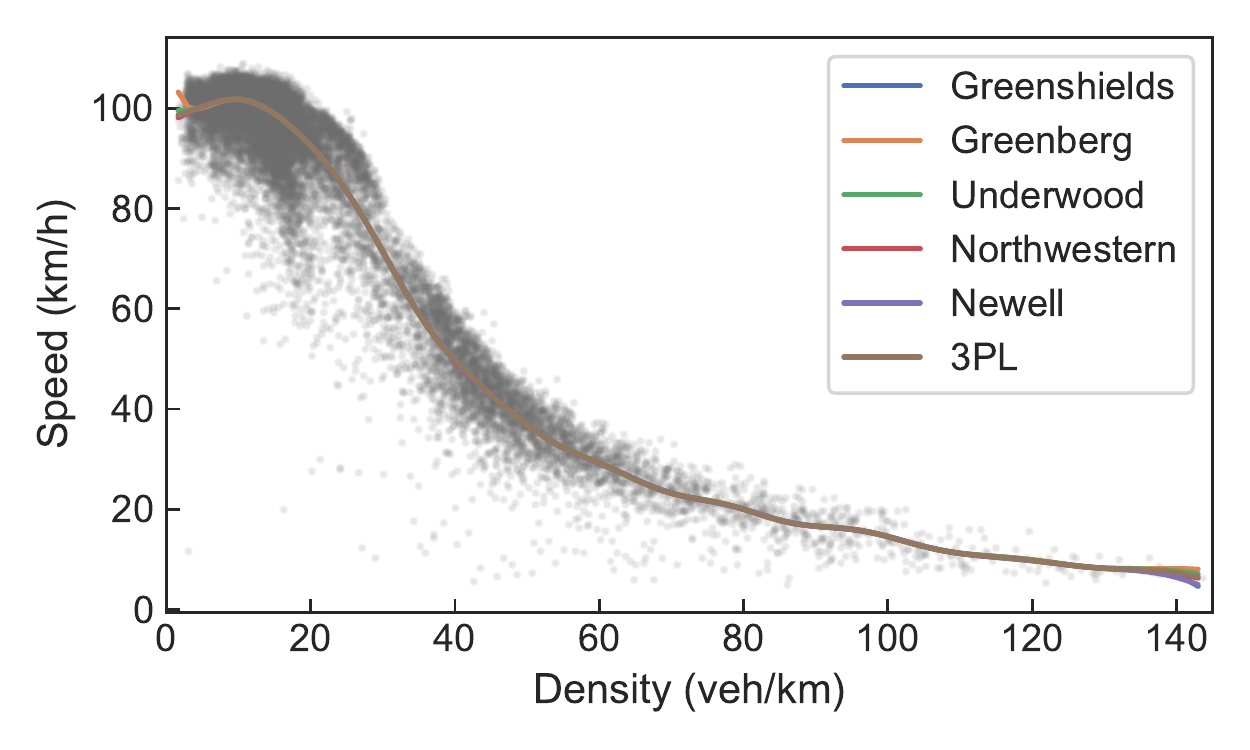}
\caption{The function values $f(k)$ fitted by the MLE of sparse GP under different speed-density mean functions.}
\label{fig:gp_values}
\end{center}
\end{figure}

The main purpose of this paper is to calibrate parameters in speed-density models, but it is also interesting to see how well a GP can fit the data. Figure~\ref{fig:gp_values} shows the function values ($f(k)$, or equivalently $m(k)+g(k)$) fitting by the MLE of GPs under different speed-density mean functions. We can see the function values are almost identical under different mean functions (except for the two edge parts). More importantly, GPs fit the empirical data much better than their mean functions, which means we can use the GP as a kind of non-parametric fundamental diagram. Although it lacks the simplicity and the mathematical properties of single-regime models, the non-parametric fundamental diagram is more close to empirical data.

Moreover, hyperparameters in the SE kernel have practical implications and can assist in diagnosing the goodness of fit of a mean function. The length scale $\ell$ determines the length of the ``waves'' in the residuals; a large $\ell$ indicates long-distance correlations in the residuals, and a small $\ell$ means locally correlated residuals. The kernel variance $\sigma^2$ determines the average distance of the function $f$ away from its mean; a smaller $\sigma^2$ indicates better goodness of fit. Table~\ref{tab:hyperparameters} shows the hyperparameters estimated by the MLE and the MCMC method, where the values of the MCMC method are sample means. The Greenshields and the Greenberg model have the largest kernel variance and length scale, indicating poor fitness. In contrast, the 3PL model estimated by the MCMC has the smallest kernel length scale and variance, meaning the best fitting result. The conclusions drawn from the hyperparameters are consistent with the observations in Figure~\ref{fig:compare} and Figure~\ref{fig:rmse}.

\begin{table}[htbp]
    \centering
    \caption{Kernel hyperparameters estimated by the MLE and the MCMC methods.}
    \small
      \begin{tabular}{clcccccc}
      \toprule
        Method  & Hyperparameters & Greenshields & Greenberg & Underwood & Northwestern & Newell & 3PL \\
      \midrule
      \multirow{2}[2]{*}{MLE} & Kernel length scale $\ell$ & 11.27 & 11.16 & 9.42  & 10.36 & 10.31 & 10.12 \\
            &                   Kernel variance $\sigma^2$ & 84.26 & 89.32 & 24.62 & 34.01 & 24.11 & 38.06 \\
      \midrule
      \multirow{2}[2]{*}{MCMC} & Kernel length scale $\ell$ & 12.23 & 12.97 & 10.37 & 10.85 & 10.75 & 9.91 \\
            &  Kernel variance $\sigma^2$ & 188.54 & 363.03 & 33.72 & 34.90 & 20.14 & 19.27 \\
      \bottomrule
      \end{tabular}%
    \label{tab:hyperparameters}%
  \end{table}%

\subsection{Posteriors of estimations}\label{sec:mcmc_results}
\begin{figure}[htpb]
    \begin{center}
    \includegraphics[width=0.8\textwidth]{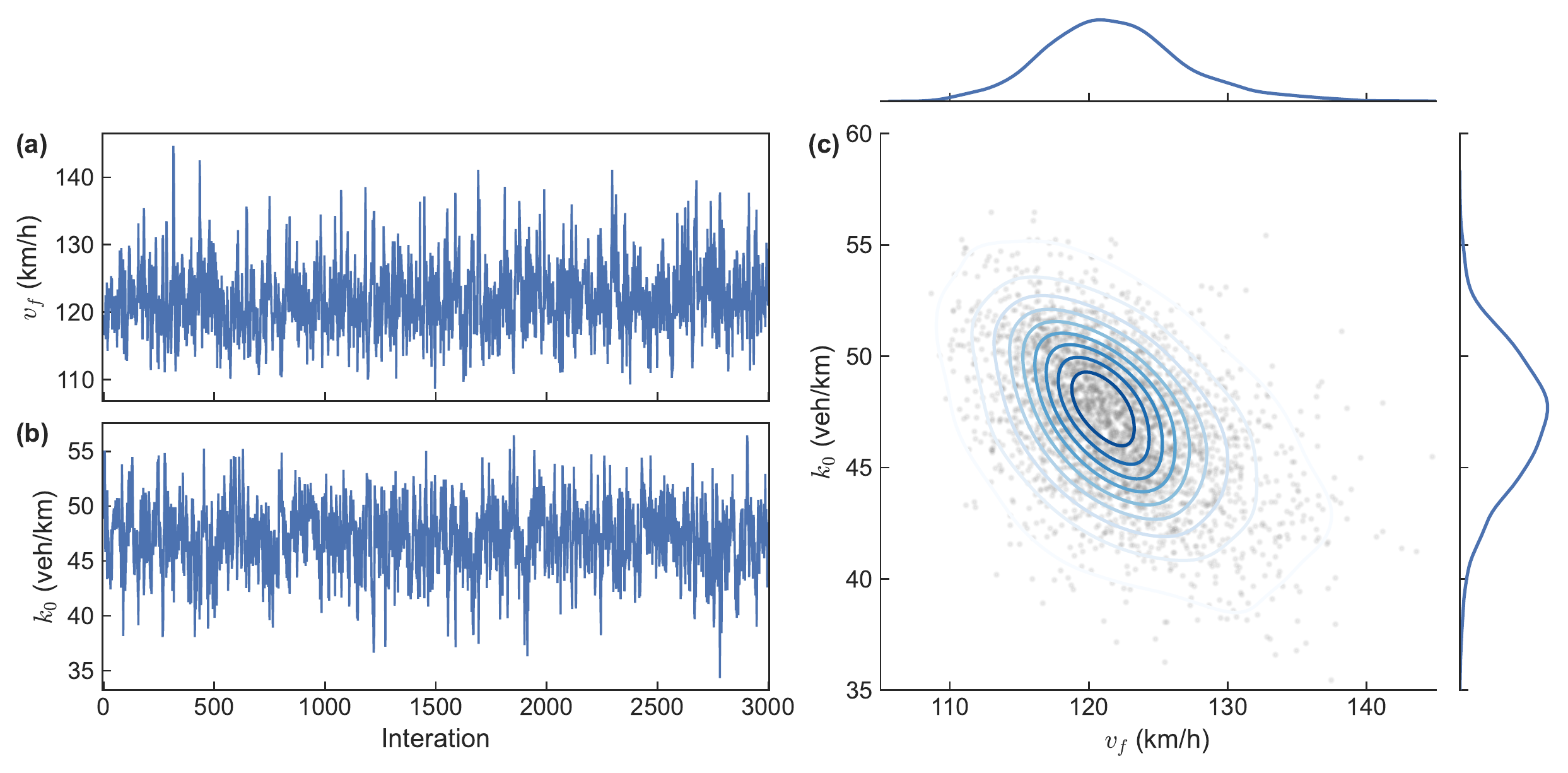}
    \caption{MCMC sampling traces and the posterior distributions of parameters for the Underwood model.}
    \label{fig:sample trace}
    \end{center}
\end{figure}

We use the Underwood model as an example to show samples drawn by the MCMC. The steady traces in Figure~\ref{fig:sample trace} (a) and (b) indicate the MCMC sampling has converged. The top and right panel in Figure~\ref{fig:sample trace} (c) are empirical distributions of $v_f$ and $k_0$, respectively. The gray dots in Figure~\ref{fig:sample trace} (c) are samples and the contour is the joint distribution of $v_f$ and $k_0$. We can find that samples of $v_f$ and $k_0$ are negatively correlated. We can also obtain the credible intervals for each parameter using the samples' distribution. The MCMC sampling helps to understand the uncertainty of the estimation.

\begin{figure}[!h]
    \begin{center}
    \includegraphics[width=0.8\textwidth]{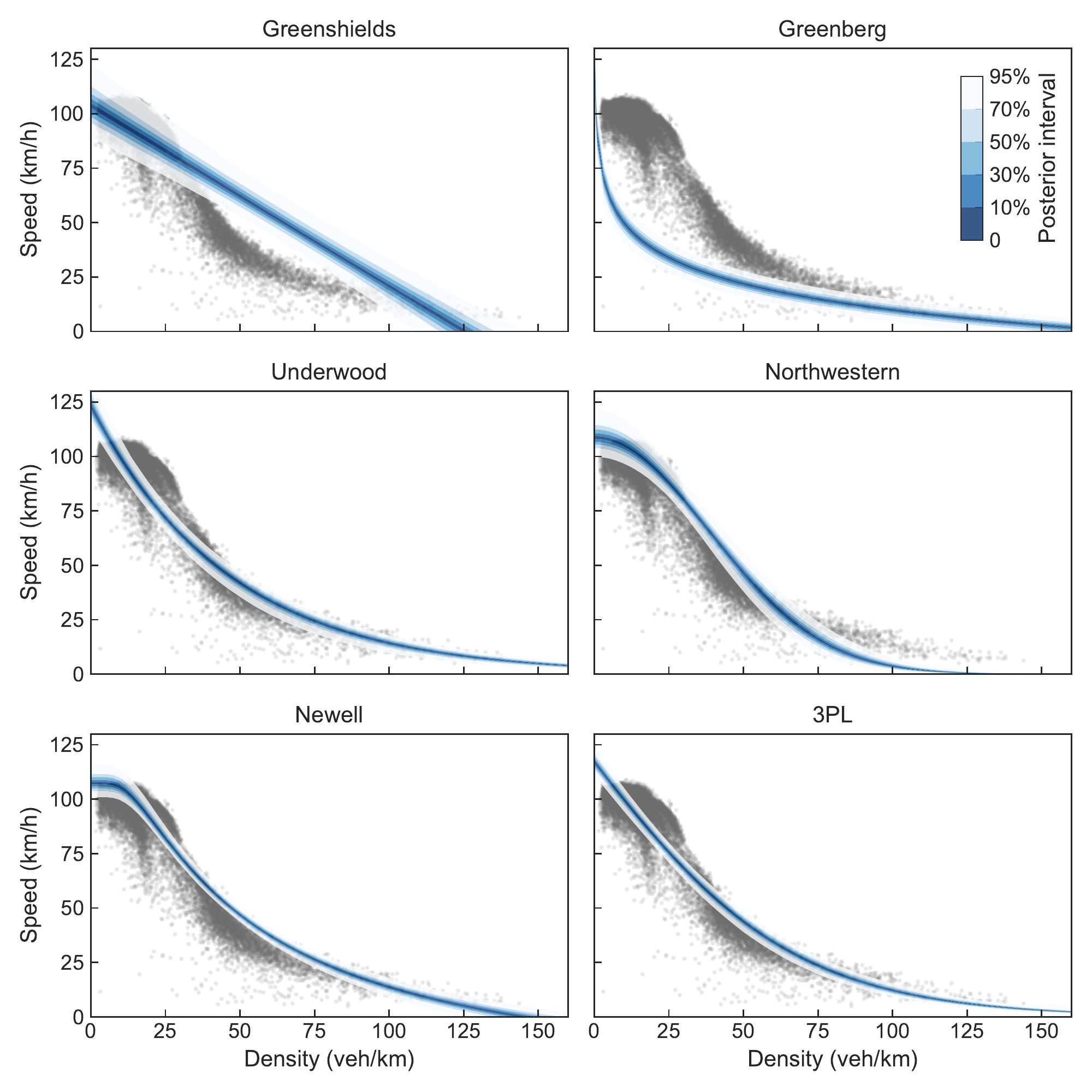}
    \caption{Posterior distributions of speed-density functions.}
    \label{fig:density plot}
    \end{center}
\end{figure}

\begin{figure}[!h]
    \begin{center}
    \includegraphics[width=0.8\textwidth]{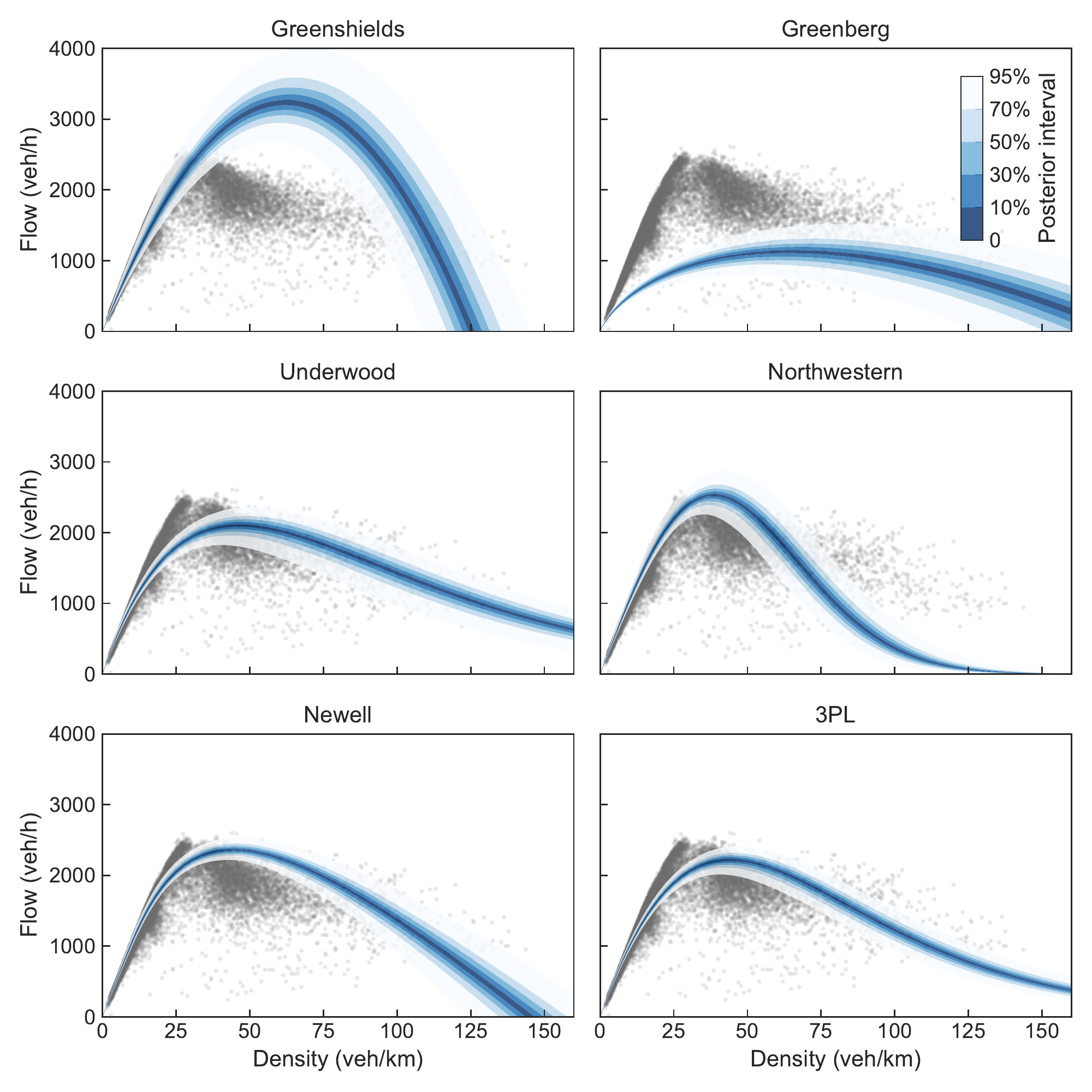}
    \caption{Posterior distributions of flow-density functions.}
    \label{fig:density plot qk}
    \end{center}
\end{figure}

A more intuitive way to examine the uncertainty in the calibration is to visualize the posterior distributions of speed-density functions. Figure~\ref{fig:density plot} shows the posterior distributions of speed-density functions by different equal-tailed intervals \citep[ETI,][]{makowski2019bayestestr}. For example, a 95\% ETI has 2.5\% of the samples on either side of this interval. We can see the posterior of the Greenshields models has the widest ETI, meaning a large uncertainty. In contrast, the Newell and the 3PL models, both with three parameters, have relatively narrow ETIs. The ETI of a model has varying widths at different density levels, such as the free flow region of the Newell model has a larger ETI than the other part of the curve. In addition to the speed-density function, we can also examine the posterior distribution of the flow-density function (obtained by $k\times v$). From Figure.~\ref{fig:density plot qk}, we can find that the estimation of 3PL flow-density function has relatively good fitness and narrow ETI. Overall, examining posterior distributions in Figure~\ref{fig:density plot} and Figure~\ref{fig:density plot qk} helps understand the calibration uncertainty and choose among different speed-density models, particularly when there are insufficient observations.

\section{Conclusions and discussions}\label{sec:conclusions}
The speed-density relationship is an essential concept in traffic flow theory. This article proposes a new low-biased calibration method for single-regime speed-density functions by modeling the residual dependencies using a zero-mean GP. Our approach based on GP has a solid statistical explanation and is a generalization of the existing WLS and LS methods. Experiments show that GP-based methods significantly reduce the biases in the LS calibration and achieve a similar effect as the WLS method. The GP can also be used as a non-parametric way to model speed-density relationships.  Moreover, we put the GP regression in a Bayesian setting, leveraging different prior distributions and noise assumptions, and using an MCMC sampling algorithm to get the posterior distributions of parameters. Besides the problem in this paper, the GP-based method can also be used to calibrate other more general problems, such as macroscopic fundamental diagrams \citep{daganzo2007urban, geroliminis2008existence} and speed-crash severity relationship \citep{aarts2006driving}. Particularly, our method is beneficial for calibration problems with correlated residuals, biased sample distributions, and when knowing parameters' uncertainties is essential (e.g., too few samples).

This research has some limitations and spaces for improvement. First, this paper only considers single-regime fundamental diagrams. Multi-regime models are usually not differentiable at the breakpoints, which would be inappropriate to use a GP with the SE kernel to model its residual correlations. Second, we assume i.i.d.\ noise but the data could be heteroscedastic. An improvement can be made by using GP with heteroscedastic noise \citep[e.g.,][]{kersting2007most}. Lastly, we can use GP with non-stationary kernels \citep{heinonen2016non} to better model the residuals in some extreme cases (e.g., the Greenberg model).


\section*{Acknowledgements}
This research is supported by the Natural Sciences  and Engineering Research Council (NSERC) of Canada and the Canada Foundation for Innovation (CFI). X. Wang would like to thank FRQNT for providing the B2X Doctoral Scholarship.

\bibliographystyle{elsarticle-harv}
\bibliography{ref}

\end{document}